\documentclass[twocolumn,prl,preprintnumbers,amsmath,amssymb]{revtex4}

\usepackage{graphicx}
\usepackage{dcolumn}
\usepackage{bm}

\begin{document}

\title{Two-Channel Kondo Effects in Al/AlO$_{x}$/Sc Planar Tunnel Junctions}

\author{Sheng-Shiuan Yeh$^{1,}$}
\email{yehshengshiuan@gmail.com}

\author{Juhn-Jong Lin$^{1,2,}$}
\email{jjlin@mail.nctu.edu.tw}

\affiliation{$^{1}$Institute of Physics, National Chiao Tung University,
Hsinchu 30010, Taiwan\\
$^{2}$Department of Electrophysics, National Chiao Tung University,
Hsinchu 30010, Taiwan}

\date{\today}

\begin{abstract}

We have measured the differential conductances $G(V,T)$ in several
Al/AlO$_{x}$/Sc planar tunnel junctions between 2 and 35 K. As the temperature
decreases to $\sim$ 16 K, the zero-bias conductance $G(0,T)$ crosses over from
a standard $-$ln$T$ dependence to a novel $- \sqrt{T}$ dependence.
Correspondingly, the finite bias conductance $G(V,T)$ reveals a two-channel
Kondo scaling behavior between $\sim$ 4 and 16 K. The observed two-channel
Kondo physics is ascribed to originating from a few localized spin-$\frac12$
Sc atoms situated slightly inside the AlO$_x$/Sc interface.

\end{abstract}

\maketitle

How conduction electrons interact with local degeneracies, which is the
central theme of the Kondo effect \cite{Hewson}, is a long-standing issue in
many-body physics. In the original Kondo effect, the local degeneracies are
provided by a spin-$\frac12$ impurity, antiferromagnetically coupled to a
single reservoir of free electrons [the single-channel Kondo (1CK) effect].
Well below a characteristic energy, the Kondo temperature $T_{K}$, the
localized moment is fully screened by the surrounding itinerant electrons to
form a singlet ground state, leading to a standard Fermi liquid behavior
\cite{Hewson}. However, Nozi\`eres and Blandin \cite{Noz-MCK} proposed that,
in the multi-channel case, i.e., the screening channels $M>2S$, where $S$ is
the spin of the localized moment, a non-Fermi liquid behavior may occur at low
temperatures. The simplest version of the multi-channel Kondo phenomena is the
two-channel Kondo (2CK) effect ($M=2$) which has recently attracted much
theoretical
\cite{Zawadowski1998,Andraka1994-Kim1995,Ott1983-Cox1987-Seaman1991,
Zawadowski1980-Muramatsu1986} and experimental
\cite{Ralph1994,Potok2007,Oreg2003,Cichorek2005} attention. Apart from a
physical spin, the local degeneracies may arise from orbital quadrupolar
degrees of freedom or nearby atomic positions, i.e., two-level systems (TLS)
\cite{Zawadowski1998}. The magnetic and quadrupolar models have been utilized
to explain the specific heat anomalies in certain heavy fermion compounds
\cite{Zawadowski1998,Andraka1994-Kim1995,Ott1983-Cox1987-Seaman1991}, while
the TLS-induced 2CK physics \cite{Zawadowski1998,Zawadowski1980-Muramatsu1986}
has recently been experimentally realized in nanoscale metal point contacts
\cite{Ralph1994}. Very lately, an artificial spin-$\frac12$ semiconductor
quantum dot coupled to two independent electron reservoirs has been elegantly
constructed \cite{Potok2007} to test the 2CK physics \cite{Oreg2003}. Besides,
the 2CK effect on electrical resistivity due to TLS is argued to be found in a
ThAsSe single crystal \cite{Cichorek2005}. Thus far, there has been no
observation of the 2CK effect caused by the ``simple" 3$d$ magnetic
transition-metal atoms. In this work, we report our finding of a non-Fermi
liquid behavior in Al/AlO$_{x}$/Sc planar tunnel junctions where a number of
spin-$\frac12$ Sc atoms are present at or slightly inside the AlO$_x$/Sc
interface \cite{Wyatt1974,Appelbaum1967}.

Our planar tunnel junctions are composed of three layers: an Al (25 nm) film
and a Sc (60 nm) film, separated by an insulating AlO$_{x}$ (1.5--2 nm)
barrier. Both the Al and Sc films were thermally evaporated, while the
AlO$_{x}$ layer was grown on the top surface of the Al film by oxygen glow
discharge \cite{Yeh-thesis}. The low-temperature resistivities of our Al (Sc)
films were typically $\rho$(4\,K) $\approx$ 13 (70) $\mu\Omega$ cm. Lock-in
techniques together with a bias circuitry were employed to measure the
differential conductance $dI/dV$ as a function of both bias voltage and
temperature. The modulation voltages used were smaller than $k_{B}T$ so that
the main smearing was due to the thermal energy. The quality of the insulating
barriers was checked according to the Rowell criteria \cite{Rowell} as well as
by measuring the $dI/dV$ curves below the superconducting transition
temperature ($\approx$ 2 K) of the Al films. At 0.25 K, a deep superconducting
gap was evidenced in all junctions, ensuring that the conduction mechanism was
governed by electron tunneling. Only the results for three representative
samples (see Table~\ref{t1}) will be discussed below. However, we stress that
very similar effects have been found in a dozen of junctions, strongly
suggesting that the observed 2CK physics is robust in the Al/AlO$_x$/Sc planar
tunnel structures.

\begin{table}
\caption{\label{t1} Relevant parameters for Al/AlO$_x$/Sc tunnel junctions.
$R_J$ is junction resistance at 300 K, $A_J$ is junction area, $T_{\rm 2CK}$
is the two-channel Kondo temperature, and $T^\ast$ is a crossover temperature
defined in the text.}
\begin{ruledtabular}
\begin{tabular}{cccccc}
Sample & $R_J$ (k$\Omega$) & $A_J$ (mm$^2$) & $T^\ast$ (K) & $T_{\rm
2CK}$\,(K)
\\ \hline
A& 2.2 & $0.5 \times 0.8$ & 4 & 72 \\
B& 1.2 & $0.5 \times 1.0$ & 5 & 64 \\
C& 5.0 & $1.0 \times 1.0$ & 6 & 56 \\
\end{tabular}
\end{ruledtabular}
\end{table}

\begin{figure}
\includegraphics[scale=0.19]{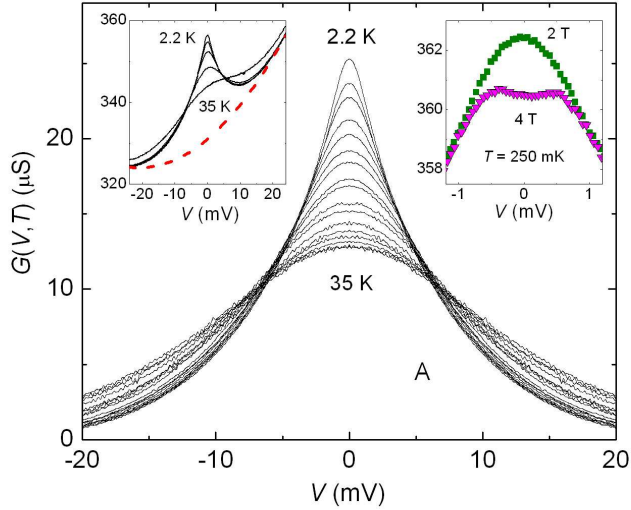}
\caption{\label{fig-Geven-H-field} (color online). $G(V,T)$ for junction A.
Left inset: Raw $dI/dV$ data. The dotted line indicates a parabolic background
conductance. Right inset: Raw $dI/dV$ data in two magnetic fields applied
normal to the plane of the junction.}
\end{figure}

The left inset to Fig.~\ref{fig-Geven-H-field} shows the raw $dI/dV$ data for
the junction A at several temperatures between 2.2 and 35 K. One clearly sees
conductance peaks around zero bias voltage, sitting on an asymmetric,
parabolic background \cite{BDR1970}. After subtracting this parabolic
background, the excess conductance $\triangle G$ contains a dominant even
contribution $G_{\rm even} \equiv \frac12[\triangle G(V) + \triangle G(-V)]$
and a minor odd contribution $G_{\rm odd} \equiv \frac12[\triangle G(V) -
\triangle G(-V)]$, where $G_{\rm odd} \lesssim 0.1 G_{\rm even}$
\cite{Yeh-thesis}. In this work, we focus on the even contribution and denote
it as $G(V,T)$. The main panel of Fig. \ref{fig-Geven-H-field} indicates that,
as $T$ reduces, the $G(V,T)$ curves become narrower and the peaks higher. Such
enhanced $G(V,T)$ cannot be expected from the disorder-induced suppression of
electronic density of states at the Fermi level due to the electron-electron
interaction effects \cite{WAL95}.

\begin{figure}
\includegraphics[scale=0.21]{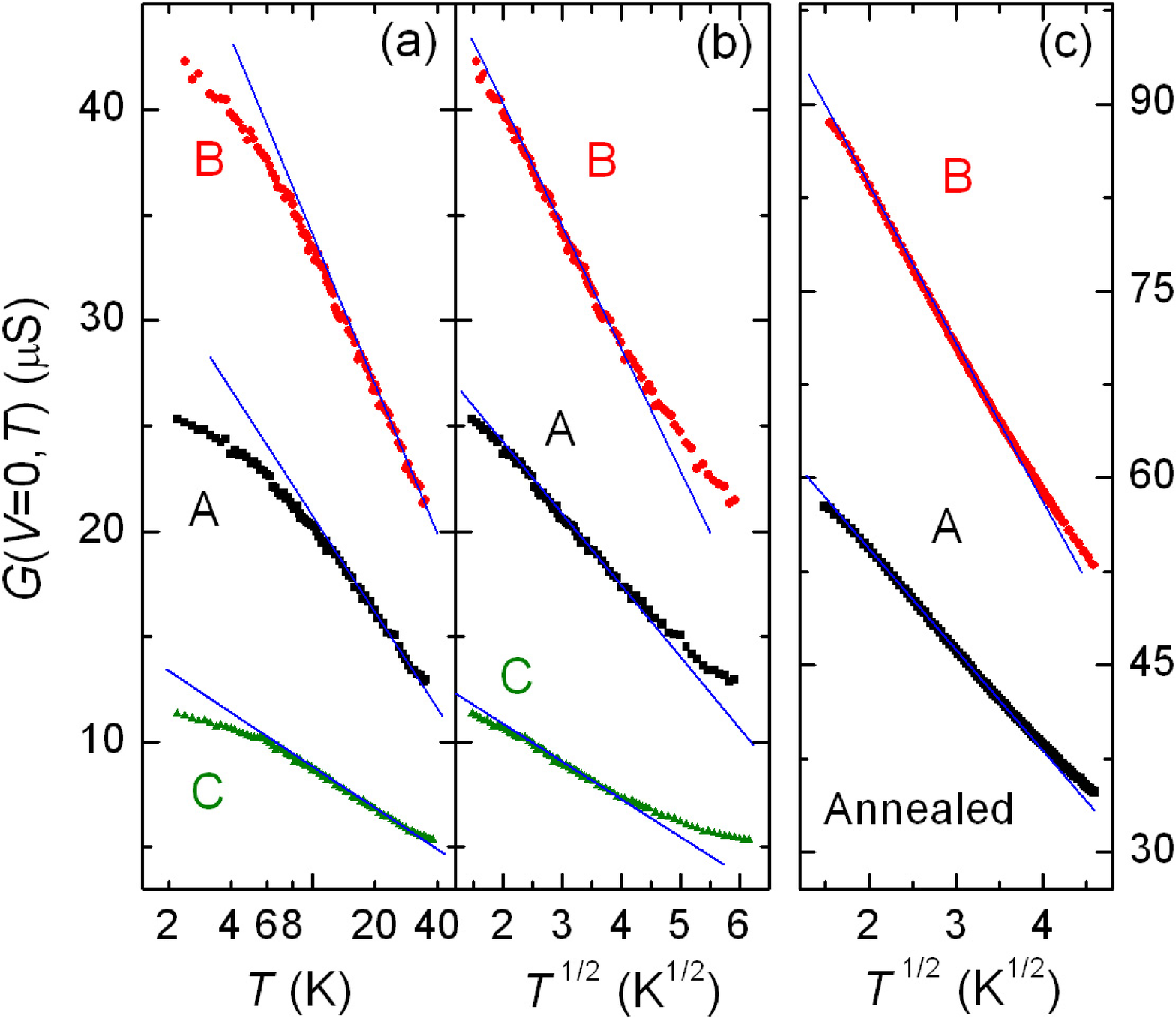}
\caption{\label{fig-G0_T} (color online). Zero-bias conductance as a function
of temperature for junctions A--C. (a) Semilog plot of $G(0,T)$ versus $T$.
(b) $G(0,T)$ versus $\sqrt{T}$ for as-prepared junctions. (c) $G(0,T)$ versus
$\sqrt{T}$ after thermal annealing.}
\end{figure}

The magnitudes of the zero-bias conductance $G(0,T)$ for junctions A--C are
plotted in Fig. \ref{fig-G0_T}. At higher temperatures ($\sim$ 16--32 K),
$G(0,T)$ obeys a $-$ln$T$ law [Fig. \ref{fig-G0_T}(a)], suggesting a
Kondo-like mechanism. Notably, in our intermediate temperature regime of
$\sim$ 5--16 K, $G(0,T)$ for all three samples obey a $- \sqrt{T}$ law [Fig.
\ref{fig-G0_T}(b)], while a deviation from the $- \sqrt{T}$ dependence starts
at about 4 K. We first notice that, in the high $T$ regime, both the $G(0,T)
\propto - {\rm ln}T$ behavior as well as our measured finite-bias $G(V,T)$
spectra \cite{Yeh-thesis} can be well described by a perturbative theory that
considers the $s$--$d$ exchange coupling between tunneling electrons and
isolated localized moments which reside slightly inside the barrier
\cite{Appelbaum1967}. This observation firmly establishes the fact that
localized moments are present in our oxide barriers. The formation of
localized moments in our junctions most likely arose from the diffusion of
some 3$d^1$ Sc atoms slightly into the AlO$_x$ barrier, e.g., during the
fabrication process \cite{Wyatt1974}.

It is known that in the 2CK state the conductance $g_{\rm 2CK}(V,T)$ due to
electrons tunneling through an individual 2CK impurity residing in the barrier
can be expressed by $g_{\rm 2CK}(V,T) = g_{\rm 2CK}(0,0) - B \sqrt{T} \Gamma
\left(AeV/k_{B}T\right)$ \cite{Affleck93,Hettler1994,Pustilnik2004}, where $A$
and $B$ are nonuniversal constants which may depend on the distance of the 2CK
impurity from the electrode/barrier interface. $\Gamma(x)$ is a universal
scaling function of $x$ with the asymptotes $1 + cx^{2}$ for $|x| \ll 1$, and
$\frac{3}{\pi}\sqrt{|x|}$ for $|x| \gg 1$, with $c \approx$ 0.0758
\cite{Affleck93,Pustilnik2004}. For macroscopic junctions, there may be a
number of 2CK impurities situating inside the barrier. If the interaction
between these 2CK impurities can be neglected, the total conductance $G_{\rm
2CK}(V,T)$ is additive: $G_{\rm 2CK}(V,T) = G_{\rm 2CK}(0,0) - \sqrt{T}
\sum_{i} B_{i} \Gamma \left( A_{i}eV/k_{B}T \right)$, where $G_{\rm 2CK}(0,0)
= \sum_{i} g_{\rm2CK}(0,0)$. Therefore, the zero-bias conductance $G_{\rm
2CK}(0,T)$ has a $-\sqrt{T}$ dependence. To eliminate $G_{\rm 2CK}(0,0)$,
which cannot be measured directly, it is very useful to scale the conductance
as a universal function of $eV/k_{B}T$: $[G_{\rm 2CK}(0,T) - G_{\rm
2CK}(V,T)]/\sqrt T = \sum_i B_i [\Gamma ( A_i eV /k_BT ) - 1 ]$. This
universal function leads to the asymptotes:
\begin{eqnarray}
\frac{G_{\rm 2CK}(0,T) - G_{\rm 2CK}(V,T)}{\sqrt T} = \Bigg\{
\begin{array}{ll}
b_{1}\left(\frac{eV}{k_{B}T}\right)^{2}\,, \mbox{ }\frac{|eV|}{k_{B}T} \ll 1 \\
b_{2}\sqrt{\frac{|eV|}{k_{B}T}}-b_{3}\,, \mbox{ }\frac{|eV|}{k_{B}T} \gg 1
\label{eq:scaling-2CK-deltaG-over-Tsquareroot}
\end{array}
\end{eqnarray}
where $b_{1}=c\sum_{i}B_{i}A_{i}^{2}$, $b_{2}=\frac{3}{\pi} \sum_{i}B_{i}
\sqrt{A_{i}}$, and $b_{3}=\sum_{i}B_{i}$. In the low bias region ($|eV|/k_{B}T
\ll 1$), the conductance can also be scaled into the form
\begin{equation}
G_{\rm 2CK}(0,T) - G_{\rm 2CK}(V,T) = f_{\rm 2CK}(T) \left( eV/k_BT
\right)^{2} \,, \label{eq:G0-2CK-low-bias}
\end{equation}
where $f_{\rm 2CK}(T)=b_{1}\sqrt{T}$.

\begin{figure}
\includegraphics[scale=0.2]{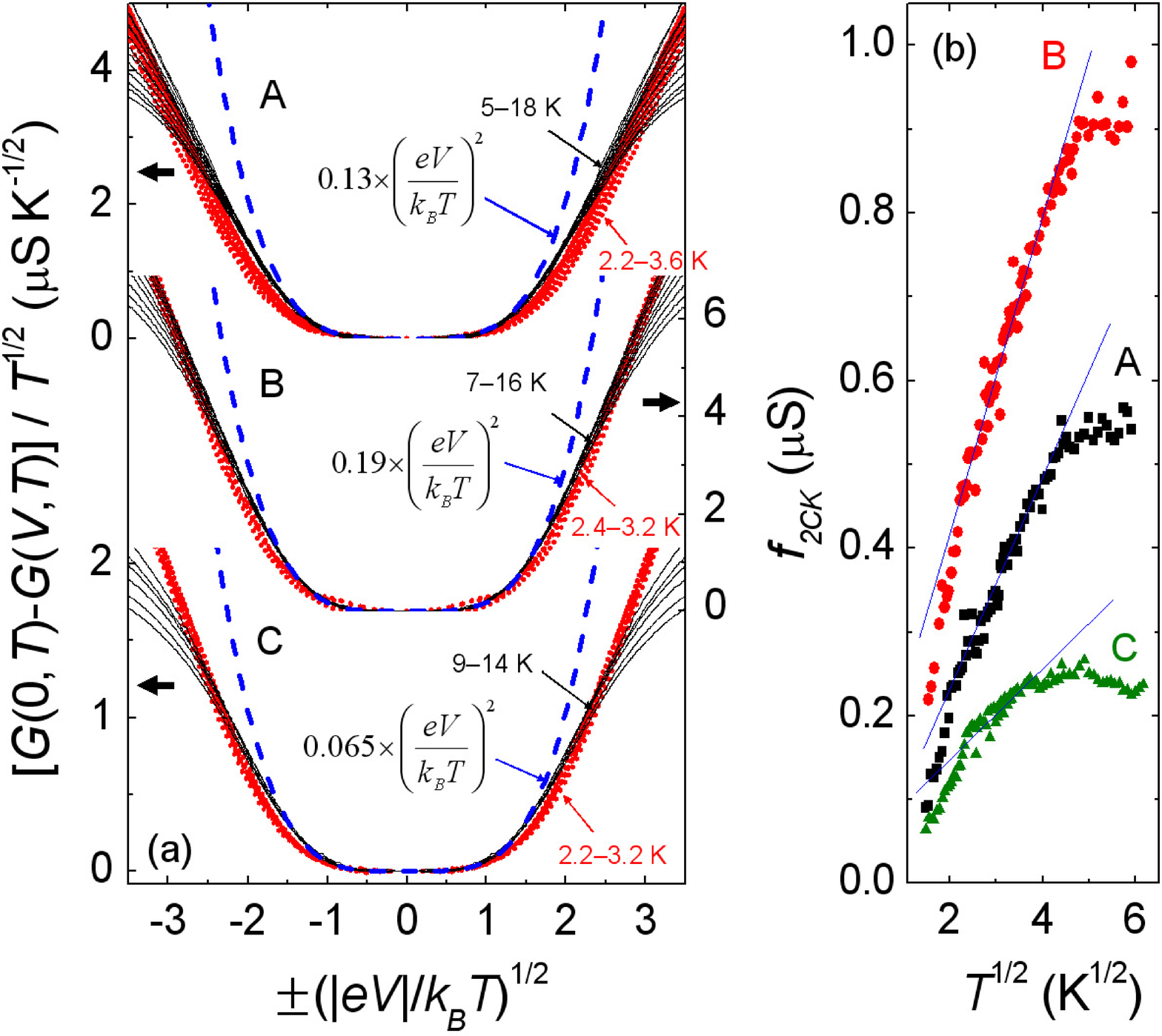}
\caption{\label{fig:2CK-scaling} (color online). (a) 2CK scaling for junctions
A--C. Solid curves stand for high $T$ data, and dotted curves for low $T$
data. (b) $f_{\rm 2CK}$ as a function of $\sqrt{T}$ for the three junctions.}
\end{figure}

The $- \sqrt{T}$ behavior of $G(0,T)$ in Fig. \ref{fig-G0_T}(b) suggests that
our junctions fall in the 2CK phase in this intermediate temperature regime.
In order to establish more and stronger evidences for this result, we have
examined the scaling behavior of the finite bias conductance $G(V,T)$. Our
measured conductances at various temperatures are scaled according to the 2CK
scaling form, Eq. (\ref{eq:scaling-2CK-deltaG-over-Tsquareroot}), and the
results are shown in Fig.~\ref{fig:2CK-scaling}(a). Notice that at
intermediate temperatures ($\sim$ 5--16 K), the data at different $T$ all
collapse onto a single curve for $(|eV|/k_BT)^{1/2} \lesssim 2.5$, with the
asymptotes correctly given by Eq.
(\ref{eq:scaling-2CK-deltaG-over-Tsquareroot}). That is, our scaled
$[G(0,T)-G(V,T)]/\sqrt T \propto (eV/k_{B}T)^{2}$ at low bias voltages of
$(|eV|/k_BT)^{1/2} \lesssim 1.4$ \cite{comment1}, while it varies linearly
with $(|eV|/k_BT)^{1/2}$ at high bias voltages. Therefore, the observed
$G(0,T) \propto - \sqrt{T}$ law together with the scaling behavior of the
$G(V,T)$ unambiguously demonstrate that our junctions fall in a 2CK state in
this intermediate temperature regime. The prefactor $f_{\rm 2CK}(T)$ in Eq.
(\ref{eq:G0-2CK-low-bias}) can be extracted from the low bias data, and is
plotted in Fig. \ref{fig:2CK-scaling}(b). It shows that $f_{\rm 2CK} \propto
\sqrt{T}$ at intermediate temperatures, but deviates at high and low
temperatures. This 2CK temperature regime for $f_{\rm 2CK}$ is in good accord
with that found in Fig.~\ref{fig-G0_T}(b) for the $G(0,T) \propto - \sqrt{T}$
attribute.

In the high temperature end, since the deviation of $G(0,T)$ from the $-
\sqrt{T}$ law starts at $\frac 14 T_{\rm 2CK}$ \cite{Hettler1994}, the
two-channel Kondo temperature, $T_{\rm 2CK}$, for our junctions may be
evaluated. In the low temperature end, there is a crossover temperature,
$T^{*}$, below which $f_{\rm 2CK}(T)$ deviates from the $- \sqrt{T}$ law. Our
values of $T_{\rm 2CK}$ and $T^\ast$ are listed in Table~\ref{t1}. It should
be noted that, in Fig.~\ref{fig:2CK-scaling}(a), deviations (red dotted lines)
from the 2CK scaling form are seen for data with $T \lesssim$ 3.6 K.

The effect of a magnetic field $H$ is shown in the right inset to Fig.
\ref{fig-Geven-H-field}. We have found $G(0,T)$ decreases with increasing $H$,
and observed a Zeeman splitting of 0.45 meV at 4 T. Such a Zeeman splitting
corresponds to a $g$-factor of 1.94, strongly indicating the presence of
localized spin-$\frac12$ moments, as discussed above. Thus, our observed
logarithmic and 2CK behaviors of $G(V,T)$ can be explained as arising from the
Sc $d$-electron impurities. The 2CK physics in our junctions should not
originate from any TLS-induced effect, since the $G(0,T) \propto -\sqrt T$
behavior in our junctions remained intact even after thermal annealing
\cite{anneal}, as is depicted in Fig. \ref{fig-G0_T}(c). In contrast, we
notice that the 2CK signals in the above-mentioned ultrasmall metal point
contacts disappeared even after annealing at room temperatures
\cite{Ralph1992}. On the other hand, the two-channel magnetic Kondo model
developed to explain the Ce-based heavy fermion compounds should require very
special symmetries in the crystal field \cite{Zawadowski1998}, which are
unlikely to happen in our junctions. Yet, another possible candidate theory is
the competition between the Kondo screening and the interimpurity interaction
$I$ proposed in the two-impurity Kondo model (2IKM)
\cite{Jones1988-1989,Zarand2006-Chung2007}, where the 2CK physics occurs at a
critical coupling strength $I_c$, which separates a Kondo-screened phase from
a local-singlet phase in the ground state. However, the existence of the 2CK
fixed point due to the 2IKM coupled to a single electron reservoir requires
some particle-hole symmetry \cite{Jones1988-1989,Zarand2006-Chung2007} which
is hard to conceive in our junctions. Thus, the microscopic origin for our
observed 2CK behavior as well as the deviation from the 2CK behavior below
about 4 K is still not well understood in terms of available theories.

On the experimental side, it is intriguing that our 2CK effect is demonstrated
in conventional planar tunnel junctions which contained 3$d^1$
transition-metal impurities. These are straightforward sample structures,
equipped with the simplest possible dynamical impurities ($S = \frac12$) for
the Kondo phenomena \cite{Hewson}. Moreover, it should be noted that a
deviation from the 2CK behavior has not been found in any previous experiments
involving more exquisite artificial structures, such as metal point contacts
\cite{Ralph1994} and semiconductor quantum dots \cite{Potok2007}, where the
characteristic Kondo temperatures are relatively low, as compared with ours.
The deviation could signify a crossover to a non-2CK phase as $T \rightarrow$
0 K. This issue deserves further investigations.

In summary, the 2CK non-Fermi liquid physics has been realized in the
differential conductances of Al/AlO$_{x}$/Sc planar tunnel junctions. In the
intermediate temperature regime, $G(0,T)$ reveals a $- \sqrt{T}$ dependence
and $G(V,T)$ obeys the 2CK scaling law. At lower temperatures, a deviation in
$G(0,T)$ from the $- \sqrt{T}$ behavior is found. These rich Kondo behaviors
are believed to originate from a few localized spin-$\frac12$ Sc atoms
situating at or slightly inside the AlO$_x$/Sc interface.

\begin{acknowledgments}

The authors thank A. M. Chang, C. H. Chung, T. A. Costi, D. L. Cox, F. Guinea,
T. K. Ng, J. Kroha, Z. S. Ma, G. Zarand, and A. Zawadowski for helpful
discussions. This work was supported by the Taiwan National Science Council
through Grant No. 96-2112-M-009-025 and by the MOE ATU Program.

\end{acknowledgments}


\begin{thebibliography}{}
\bibitem{anneal} Thermal annealing of the junctions was performed in vacuum
($\sim$ 10$^{-5}$ torr) at 100$^\circ$C for 1 h. After annealing, the
$R_J$(300\,K) of samples A and B decreased to 0.83 and 0.50 k$\Omega$,
respectively. In ploting Fig. \ref{fig-G0_T}(c), the background conductances
had been subtracted.
\end{thebibliography}
\end{document}